# Analytical coupled vibroacoustic modeling of membrane-type acoustic metamaterials: membrane model

**Yangyang Chen and Guoliang Huang** [a)]

*Department of Systems Engineering, University of Arkansas at Little Rock, Little Rock, AR 72204, USA*

**Xiaoming Zhou and Gengkai Hu**

*Key Laboratory of Dynamics and Control of Flight Vehicle, Ministry of Education, School of Aerospace Engineering, Beijing Institute of Technology, Beijing, China, 100081*

**Chin-Teh Sun**

*School of Aeronautics and Astronautics, Purdue University, W. Lafayette, IN 47907, USA*

Membrane-type Acoustic Metamaterials (MAMs) have demonstrated unusual capacity in controlling low-frequency sound transmission/reflection. In this paper, an analytical vibroacoustic membrane model is developed to study sound transmission behavior of the MAM under a normal incidence. The MAM is composed of a prestretched elastic membrane with attached rigid masses. To accurately capture finite-dimension rigid mass effects on the membrane deformation, the point matching approach is adopted by applying a set of distributed point forces along the interfacial boundary between masses and the membrane. The accuracy and capability of the theoretical model is verified through the comparison with the finite element method. In particular, microstructure effects such as weight, size and eccentricity of the attached mass, pretension and thickness of the membrane on the resulting transmission peak and dip frequencies of the MAM are quantitatively investigated. New peak and dip frequencies are found for the MAM with one and multiple eccentric attached masses. The developed model can be served as an efficient tool for design of such membrane-type metamaterials.

---

[a)] Electronic mail: glhuang@ualr.edu

## I. INTRODUCTION

The problem of a membrane subject to dynamic mechanical and/or acoustic loads is of great importance in many engineering branches such as acoustic waveguides, sound and musical instruments and vibration and noise control. This problem is revitalized by the recent experimental finding in which a membrane with attached masses can achieve a high sound insulation in low frequency ranges.[1,2] The theoretical problem of sound transmission through membranes and panels has been investigated intensively for decades.[3-5] For example, an analytic solution in an integral form of the problem of an ideal stretched circular membrane under a plane sound wave has been derived.[6] To gain better acoustic properties and increase membranes' potential energy, the membrane carrying attached masses of finite area was suggested and analyzed.[7-8] It was realized that the added small mass loadings can significantly affect vibration behaviors and acoustic transmission properties of the membrane,[9] which provided many interesting potential applications in sound or noise control schemes.[10] However, compared with vibroacoustic studies of the pure membranes, studies on membranes with distributed masses are not well established because of the difficulty from its complex multiconnected boundary conditions.

The prototype proposed by Yang et al. now is recast into more general concept of membrane-type metametrials,[1,2] which have been suggested to possess excellent acoustic properties for low-frequencies sound insulation. Compared with the conventional sound attenuation materials typically utilizing thermally-coupled dissipation mechanisms and suffering from inadequate low frequency sound attenuation, the MAMs can be designed to possess nearly total reflection for targeting low-frequency acoustic sources. The basic microstructure of this MAM consists of a prestreched membrane with one or many attached small heterogeneous masses acting as resonators. Fixed outer edges are imposed by a relatively rigid grid. The unusual low-frequency vibroacoustic behavior was numerically characterized, and the response spectrum shows separate transmission peaks at resonances. The low-frequency sound transmission mechanism was also numerically explained through effective mass density and averaged normal displacement by using the finite element method, although the nearly total reflection of the MAM is of limited frequency bandwidth. To address this issue, many attempts on the MAM microstructure designs have been made in producing multiple transmission loss peaks including the number, distribution and geometries of attached rigid masses.[11-13] The designed MAMs were validated by conducting experimental testing to measure transmission losses of samples. However, for optimized microstructure design and better characterization of the MAM, a solid analytical model to accurately capture its vibroacoustic behavior of the attached masses is highly needed. The analytical method can also be very useful in the initial design of MAMs for desired engineering applications. The advantages of the analytical model are its computational convenience and flexibility, which enables it to reveal the physical insight of the MAM. A simplified theoretical study has been conducted to analyze the vibroacoustic behavior of the MAM with one attached mass.[14] The model can give reasonable prediction on the transmission loss peak and valleys caused by the system's antiresonance and resonance motion. However, structural-acoustic couplings and the attached mass motion are not properly explored.

In this paper, we develop a new vibroacoustic model to investigate dynamic behavior of the MAM with one and multiple attached mass resonators. The general motion of the rigid attached masses in finite domains is properly described by using both translational and rotational degrees of freedoms. The point matching approach is adopted to seek eigenmodes of the MAM, in which interactions of the finite rigid masses upon the deformable membrane are represented by a set of distributed point forces along their boundaries. By solving a vibroacoustic integrodifferential equation, local deformation field in the membrane can be determined by using the modal superposition method. What distinguishes this study from others is that microstructure effects, such as weight, size and eccentricity of the attached mass, pretension and thickness of the membrane on the transmission peak and dip frequencies of the MAM can be quantitatively identified. It is hoped that such solutions may lead to general conclusions that will be of help to the future MAM designer.

## II. THEORETICAL MEMBRANE MODEL

Consider now a unit cell of a MAM, as shown in Fig. 1(a), where the deformable membrane is attached by an eccentrically circular mass in finite dimension. In the figure, the membrane is subject to initial uniform tension $T$ per unit length. The radius and density per unit area of the membrane are denoted as $R$ and $\rho_s$, and the radius and density per unit area of the mass are $a$ and

$\rho_c$. In the figure, $d$ represents eccentricity of the mass in the Cartesian coordinate system $(x,y)$ with origin $O$ in the center of the membrane. The polar coordinate $(r,\theta)$ is also taken in the figure. Another local Cartesian coordinate system $(x', y')$ and its corresponding polar coordinate system $(r',\theta')$ are established with origin $O'$ in the center of the mass. The inner boundary between the mass and the membrane is denoted as some discrete points $(x_i,y_i)$ in the Cartesian coordinate system $(x,y)$, $(b_i, \Theta_i)$ in the global polar coordinate system $(r,\theta)$ and $(a,\Theta_i')$ in the local polar coordinate system $(r',\theta')$. In the study, we focus on the sound transmission and reflection of the stretched MAM in a tube subject to a plane normal sound wave, as shown in Fig. 1(b). Perfectly absorbing boundary conditions are assumed in both ends of the tube so that there will be no multiple reflected waves to the MAM. In the study, the free vibration problem of the MAM involving with the multi-connected boundaries such as the fixed circular outer and displacement continuous eccentric inner boundaries is first studied by using the point matching approach [Nagaya and Poltorak, 1989]. Then, the vibroacoustic coupling behavior of the MAM will be analyzed through the modal superposition theory, from which the sound transmission and reflection of the MAM can be analytically determined. This model can also be easily extended to analyze the acoustic behavior of the MAM with multiple attached masses in arbitrary shapes.

## A. Eigenvalue problem of the MAM

For the MAM, the attached mass is assumed to be perfectly bonded to the membrane and rigid compared with the deformable membrane. To properly capture effects of the finite mass on the small deformation of the membrane, the point matching scheme is applied by using distributed point forces along the interfacial boundary between the mass and the membrane. Therefore, the governing equation of the membrane can be written as

$$\rho_s \frac{\partial^2 W(x,y,t)}{\partial^2 t} - T\nabla^2 W(x,y,t) = \sum_{i=1}^{I} Q_i(t)\delta(x-x_i)\delta(y-y_i) \,, \qquad (1)$$

where $W(x,y,t)$ is the out-of-plane displacement of the membrane in the $z$ direction, $\nabla^2 = \frac{\partial^2}{\partial x^2} + \frac{\partial^2}{\partial y^2}$ or $\nabla^2 = \frac{1}{r}\frac{\partial}{\partial r}\left(r\frac{\partial}{\partial r}\right) + \frac{1}{r^2}\frac{\partial^2}{\partial \theta^2}$ in the polar coordinate, and the right hand side comes from the summation of distributed point forces at $I$ collocation points along the inner boundary of the membrane, $\delta$ is the Dirac delta function. Since only the steady-state response field will be considered, the time factor $e^{i\omega t}$, which applies to all the field variables, will be suppressed in the paper. Then, Eq. (1) in the polar coordinate can be rewritten as

$$\alpha^2 W(r,\theta) + \frac{\partial^2 W(r,\theta)}{\partial r^2} + \frac{1}{r}\frac{\partial W(r,\theta)}{\partial r} + \frac{1}{r^2}\frac{\partial^2 W(r,\theta)}{\partial \theta^2} = -\sum_{i=1}^{I} N_i \delta(r - b_i)\delta(\theta - \Theta_i), \qquad (2)$$

in which $\alpha^2 = \frac{\rho_s \omega^2}{T}$ and $N_i = \frac{Q_i}{b_i T}$. The coordinate relations for collocation points are $b_i = \sqrt{a^2 + 2ad\cos(\Theta_i') + d^2}$ and $\Theta_i = \cos^{-1}\left[\frac{d + a\cos(\Theta_i')}{b_i}\right]$.

The exact solution of Eq. (2) can be obtained by using Lagrange's method and performing integration as [Nagaya and Poltorak, 1989]

$$W(r,\theta) = \sum_{n=0}^{\infty} \epsilon_n \{A_{1n} J_n(\alpha r) - \frac{1}{2}\sum_{i=1}^{I} N_i b_i [Y_n(\alpha r)J_n(\alpha b_i) - J_n(\alpha r)Y_n(\alpha b_i)]u(r - b_i)\cos(n\Theta_i)\}\cos(n\theta) + \sum_{n=0}^{\infty} \epsilon_n \{A_{2n} J_n(\alpha r) - \frac{1}{2}\sum_{i=1}^{I} N_i^* b_i [Y_n(\alpha r)J_n(\alpha b_i) - J_n(\alpha r)Y_n(\alpha b_i)]u(r - b_i)\sin(n\Theta_i)\}\sin(n\theta), \qquad (3)$$

where $A_{1n}$ and $A_{2n}$ are arbitrary constants related to symmetric and antisymmetric modes with respect to the $x$ axis, $J_n(\alpha r)$ and $Y_n(\alpha r)$ are Bessel functions of first and second kind of order $n$, respectively, $\epsilon_n = \begin{cases} \frac{1}{2}, & \text{for } n = 0 \\ 1, & \text{for } n \geq 1 \end{cases}$, and $u(r - b_i) = \begin{cases} 1, & \text{for } r > b_i \\ 0, & \text{for } r < b_i \end{cases}$. $N_i$ and $N_i^*$ are two unknown constants related to point loadings for symmetric and antisymmetric modes.

Because the outer boundary of the membrane is fixed, we have

$$W(R,\theta) = 0. \qquad (4)$$

Substituting Eq. (3) into Eq. (4) and solving for $A_{1n}$ and $A_{2n}$, the displacement field of the membrane can be obtained as

$$W(r,\theta) = \sum_{i=1}^{I} N_i F_i(r,\theta) + N_i F_i^*(r,\theta), \qquad (5)$$

where $F_i(r,\theta) = \sum_{n=0}^{\infty} b_i \epsilon_n \left\{ \frac{J_n(\alpha r)}{2J_n(\alpha R)}[Y_n(\alpha R)J_n(\alpha b_i) - J_n(\alpha R)Y_n(\alpha b_i)] \cos(n\Theta_i) - \frac{1}{2}[Y_n(\alpha r)J_n(\alpha b_i) - J_n(\alpha r)Y_n(\alpha b_i)]u(r - b_i)\cos(n\Theta_i) \right\}\cos(n\theta)$, and $F_i^*(r,\theta) =$

$\sum_{n=0}^{\infty} b_i \epsilon_n \left\{ \frac{J_n(\alpha r)}{2 J_n(\alpha R)} [Y_n(\alpha R) J_n(\alpha b_i) - J_n(\alpha R) Y_n(\alpha b_i)] \sin(n\Theta_i) - \frac{1}{2} [Y_n(\alpha r) J_n(\alpha b_i) - \right.$

$\left. J_n(\alpha r) Y_n(\alpha b_i)] u(r - b_i) \sin(n\Theta_i) \right\} \sin(n\theta).$

The unknown constants $N_i$ and $N_i^*$ can be determined through the inner boundary condition of the membrane.

For the attached mass in finite dimension, the general rigid motion can be completely described by the translation and rotation in the local polar coordinate system $(r', \theta')$ as

$$W'(r', \theta') = \bar{a} r' \cos(\theta') + \bar{b} r' \sin(\theta') + \bar{c}, \tag{6}$$

where $\bar{a}$, $\bar{b}$ and $\bar{c}$ are three unknown constants.

Equations of the general motion of the circular mass including the translation and rotation can be written as

$$-\sum_{i=1}^{I} T N_i b_i = m \frac{\partial^2 W'}{\partial t^2}\bigg|_{(r'=0)} = -m\omega^2 \bar{c}, \tag{7}$$

$$-\sum_{i=1}^{I} T N_i b_i a \cos(\Theta_i') = I_{y'} \frac{\partial^2 \psi_{y'}'}{\partial t^2} = -I_{y'}\omega^2 \bar{a}, \tag{8}$$

$$-\sum_{i=1}^{I} T N_i^* b_i a \sin(\Theta_i') = I_{x'} \frac{\partial^2 \psi_{x'}'}{\partial t^2} = -I_{x'}\omega^2 \bar{b}, \tag{9}$$

where $m$, $\psi_{y'}'$, $\psi_{x'}'$, $I_{y'}$ and $I_{x'}$ are the weight of the mass, rotational displacement with respect to the local $y'$ axis, rotational displacement with respect to the local $x'$ axis, the moment of inertia with respect to the local $y'$ axis and the moment of inertia with respect to the local $x'$ axis, respectively. From Eqs. (7) - (9), the unknown constants $\bar{a}$, $\bar{b}$ and $\bar{c}$ can be obtained in terms of $N$ and $N^*$.

Finally, based on the displacement continuity between the mass and the membrane on every collocation point, we have

$$W'(a, \Theta_j') - W(b_j, \Theta_j) = 0. \tag{10}$$

Substituting Eqs. (5) - (9) into Eq. (10) yields

$$\sum_{i=1}^{I} N_i \left( \bar{a}_i a \cos(\Theta'_j) + \bar{c}_i - F_i(b_j, \Theta_j) \right) + N_i^* \left( \bar{b}_i^* a \sin(\Theta'_j) - F_i^*(b_j, \Theta_j) \right) = 0. \qquad (11)$$

where $\bar{c}_i = \frac{Tb_i}{m\omega^2}$, $\bar{a}_i = \frac{Tb_i a \cos(\Theta'_i)}{I_{y'} \omega^2}$ and $\bar{b}_i^* = \frac{Tb_i a \sin(\Theta'_i)}{I_{x'} \omega^2}$.

The eigenvalues of the Eq. (11) are obtained by setting the determinant of the coefficient matrix in the above system of equations for $\{N_i, N_i^*\}$ of size $(2I \times 2I)$ to be zero, from which the wave mode shapes of the membrane can be also obtained.

## B. Vibroacoustic modeling of the MAM

Consider a plane sound wave is normally incident on the MAM. The objective is to determine the pressure ratio between the incident and transmitted sound waves. To consider the vibroacoustic coupling behavior, the governing equation of the acoustic excited membrane with an attached mass can be expressed as

$$-\omega^2 \rho_s W(x,y) - T\nabla^2 W(x,y) = p_1|_{(z=0)} - p_2|_{(z=0)} + \sum_{i=1}^{I} Q_i \ \delta(x - x_i)\delta(y - y_i), \qquad (12)$$

where $p_1$ and $p_2$ are sound pressures in the left and right sides of the membrane, which can be decomposed as

$$p_1(r, \theta, z) = p_I + p_R + p^- = P_I e^{-ik_1 z} + P_R e^{ik_1 z} + p^-(r, \theta, z), \qquad (13)$$

$$p_2(r, \theta, z) = p_T + p^+ = P_T e^{-ik_1 z} + p^+(r, \theta, z), \qquad (14)$$

where $p_I$, $p_R$ and $p_T$ are incident, reflected and transmitted plane waves with complex amplitudes being $P_I$, $P_R$ and $P_T$, respectively, $p^-$ and $p^+$ denote higher order wave fields scattered by the membrane, and $k_1$ is the acoustic wave number in the air.

Out-of-plane displacement field can also be decomposed as

$$w(x,y) = \langle w \rangle + \delta w, \qquad (15)$$

in which $\langle \cdot \rangle$ denotes the average of the parameter.

Air particles and MAM's out-of-plane velocities must be continuous on MAM's surfaces,

$$\frac{\partial p_1}{\partial z}\bigg|_{(z=0^-)} = \frac{\partial p_2}{\partial z}\bigg|_{(z=0^+)} = \omega^2 \rho_1 W, \qquad (16)$$

where $\rho_1$ is the mass density of air.

Substituting Eqs. (13) - (15) into (16), we obtain

$$\frac{\partial(p_I+p_R)}{\partial z}\Big|_{(z=0^-)} + \frac{\partial p^-}{\partial z}\Big|_{(z=0^-)} = \frac{\partial p_T}{\partial z}\Big|_{(z=0^+)} + \frac{\partial p^+}{\partial z}\Big|_{(z=0^+)} = \omega^2\rho_1\langle w\rangle + \omega^2\rho_1\delta w. \tag{17}$$

We may assume that far field transmitted and reflected plane waves in the tube, $p_R$ and $p_T$, are produced by the piston-like motion of the MAM, and higher order scattered fields, $p^-$ and $p^+$, are caused by the $\delta w$.

Velocity continuity for the 0-th order plane waves is

$$\frac{\partial(p_I+p_R)}{\partial z}\Big|_{(z=0^-)} = \frac{\partial p_T}{\partial z}\Big|_{(z=0^+)} = \omega^2\rho_1\langle w\rangle, \tag{18}$$

and velocity continuity for higher order scattered waves is

$$\frac{\partial p^-}{\partial z}\Big|_{(z=0^-)} = \frac{\partial p^+}{\partial z}\Big|_{(z=0^+)} = \omega^2\rho_1\delta w. \tag{19}$$

The higher order scattered waves from the membrane should satisfy

$$p^-|_{(z=0)} = -p^+|_{(z=0)}, \tag{20}$$

and define $p^- = P$.

Equations (18)-(20) give

$$P_T = i\omega\rho_1 c_1\langle W\rangle, \tag{21}$$

$$P_R = P_I - P_T, \tag{22}$$

$$\frac{\partial P}{\partial z}\Big|_{(z=0)} = \omega^2\rho_1\delta w, \tag{23}$$

$$p_1|_{(z=0)} - p_2|_{(z=0)} = 2\big(P_I - P_T + P(r,\theta,0)\big), \tag{24}$$

where $c_1$ is the sound velocity in air.

The higher order scattered acoustic wave field $P(r,\theta,z)$ is governed by

$$\Delta_0 P(r,\theta,z) + k_1^2 P(r,\theta,z) = 0, \tag{25}$$

with the rigid boundary condition along the inner surface of the tube as

$$\left.\frac{\partial P}{\partial r}\right|_{(r=R)} = 0, \tag{26}$$

and $\Delta_0$ being the three-dimensional Laplacian.

Applying the appropriate divergence theorem with Green's function, we obtain

$$P(r,\theta,0) = \int_{-\pi}^{\pi}\int_0^R G(r,\theta,0,r^*,\theta^*,0)\left.\frac{\partial P(r^*,\theta^*,z^*)}{\partial z^*}\right|_{(z^*=0)} r^*dr^*d\theta^* =$$

$$\omega^2 \rho_1 \int_{-\pi}^{\pi}\int_0^R G\delta w r^*dr^*d\theta^*, \tag{27}$$

where the Green's function of the acoustic wave field is

$$G = \frac{e^{ik_1 S}}{4\pi S} + \frac{e^{ik_1 S_1}}{4\pi S_1}, \tag{28}$$

with the boundary condition being $\left.\frac{\partial G}{\partial z}\right|_{(z=0)} = 0$ and

$$S = \sqrt{r^2 + r^{*2} - 2rr^*\cos(\theta - \theta^*) + (z - z^*)^2},$$

$$S_1 = \sqrt{r^2 + r^{*2} - 2rr^*\cos(\theta - \theta^*) + (z + z^*)^2}.$$

Combining Eqs. (12), (21), (24) and (27), the governing equation of coupled vibroacoustic modeling of the MAM can be obtained as

$$-\rho_s\omega^2 W - T\nabla^2 W + 2i\omega\rho_1 c_1\langle W\rangle - 2\omega^2\rho_1\int_{-\pi}^{\pi}\int_0^R G(W - \langle W\rangle)r^*dr^*d\theta^* = 2P_I +$$

$$\sum_{i=1}^I Q_i\ \delta(r - b_i)\delta(\theta - \Theta_i), \tag{29}$$

in which the integral operator represents the sound pressure on the membrane due to the membrane's motion.

To solve the integrodifferential equation, we use the modal superposition method such that the motion of the membrane is assumed to be

$$W(x,y) = \sum_{k=1}^{+\infty} W_k(x,y)q_k = \sum_{k=1}^{+\infty} W_k(r,\theta)q_k, \tag{30}$$

where $q_k$ is the unknown constant to be determined, and $W_k$ is the $k$-th order mode shape, which satisfies

$$\rho_s \omega_k^2 W_k + T\nabla^2 W_k = -T\sum_{i=1}^{l} N_i^{(k)} \delta(r - b_i)\delta(\theta - \Theta_i), \qquad (31)$$

with $\omega_k$ and $N_i^{(k)}$ being the $k$-th order natural frequency and dimensionless constants corresponding to $W_k$. By substituting Eq. (30) into Eq. (29), multiplying each term with $W_l$ and integrating over the whole area of the MAM, combining Eqs. (7) - (9) and Eq. (31), and considering the orthogonality of eigenfunctions , we have

$$(\omega_l^2 - \omega^2)\left(\rho_s \pi R^2 \langle W_l^2 \rangle + m\bar{c}_l^2 + I_{y'}\bar{a}_l^2 + I_{x'}\bar{b}_l^2\right)q_l + \sum_{k=1}^{+\infty} 2i\omega\rho_1 c_1 \pi R^2 \langle W_l\rangle\langle W_k\rangle q_k -$$
$$\sum_{k=1}^{+\infty} 2\omega^2\rho_1 \int_{-\pi}^{\pi}\int_0^R W_l \int_{-\pi}^{\pi}\int_0^R G(W_k - \langle W_k\rangle)r^* dr^* d\theta^* r dr d\theta \cdot q_k = 2P_l\pi R^2\langle W_l\rangle, \qquad (32)$$

in which $\bar{a}_l$, $\bar{b}_l$ and $\bar{c}_l$ are $\bar{a}$, $\bar{b}$ and $\bar{c}$ corresponding to $W_l$, respectively. The unknown constant $q_k$ can be determined by solving the above linear equations.

The average velocity of the MAM can be calculated as

$$i\omega\langle W\rangle = i\omega\sum_{k=1}^{+\infty}\langle W_k\rangle q_k. \qquad (33)$$

Therefore, the acoustic transmission coefficient can be defined as

$$\tilde{T} = \frac{P_T}{P_I} = \frac{i\omega\rho_1 c_1\langle W\rangle}{P_I}, \qquad (34)$$

and the intensity transmission coefficient is

$$T_I = \left|\tilde{T}\right|^2. \qquad (35)$$

## III. VALIDATION OF THE THEORETICAL MODELING

To evaluate the developed model, acoustic property predictions of the MAM from the current model will be compared with those by using commercially available finite element (FE) code, COMSOL Multiphysics. In the FE modeling, the three-dimensional solid element is selected for the membrane and the attached mass. To simulate acoustic wave transmission and reflection of the MAM, shown in Fig. 1(b), a normal incident acoustic wave is applied on the left of the tube. Multiple physical boundary continuities, such as face loadings on the solid and the acceleration within acoustic field, along the interface between the membrane and the air are constructed. A perfectly clamped boundary condition is defined on the outer edge of the MAM, and the rigid wall boundary condition is assumed on the side boundary of the cylinder tube. Two acoustic radiation boundaries are assumed on the both ends of the system. Material properties and pretension of the membrane and parameters of the attached mass are given in Table 1. For the current analytical model, the expansions in Eq. (11) are truncated to the collocation points $I = 15$ on the half of the inner circular boundary for the solution convergence, when geometrically symmetric properties are considered.

The three lowest natural frequencies of the symmetric modes of MAMs obtained from the analytical method and finite element analysis with both the central mass ($d = 0$ mm) and the eccentric mass ($d = 6$ mm) are listed in Table 2. Very good agreement is observed for all the first three resonant frequencies, which confirms the accuracy and capability of the current analytical model. The corresponding mode shapes from the analytical method and finite element analysis are also shown in Fig. 2. It can be found that the analytical model can provide almost the same predictions as those from the FE method. In particular, it is also very interesting to notice that the prediction from the current model can accurately capture geometric effects of the attached mass on the deformable membrane through the general mass motion. For the MAM with the attached central mass, the first deformation mode is caused by the translational motion of the attached mass, and the second and third mode shapes are caused by the large vibration of the membrane region with the central mass being motionless. However, for the MAM with the eccentric attached mass, the first deformation mode is caused by both translational and rotational motions of the mass, the second mode shape is mainly caused by the rotational motion of the mass with almost no translational motion and the third mode shape is caused by the large deformation of the membrane.

Figure 3 illustrates the obtained intensity transmission coefficients of the MAM from the FE simulation (solid curve) and analytical results (dash curve) with a centric attached mass ($d = 0$ mm). It is observed that the acoustic behavior of the MAM does not follow the prediction by the conventional mass density law, two transmission peaks are observed in the two lowest resonant frequencies and there is a dip frequency between them. Based on the mode shapes shown in Fig. 2, it is noticed that the first transmission peak is caused by the eigenmode in which the membrane and the attached mass vibrate in phase, while the second peak is only due to the membrane's motion. To interpret the physical mechanism of the MAM, the dynamic effective mass density of the MAM is also calculated and plotted in Fig. 3 (dot curve), which is defined as $\rho_{\text{eff}} = \langle p_1 - p_2 \rangle / \langle a_z \rangle$. Starting from the first peak frequency, the effective mass density turns from positive to negative, and approaches to negative infinite values around the dip frequency, then jumps to positive infinite values, and gradually decreases to zero at the second peak frequency. The effective mass density of the MAM can be also explained by the displacement field of the membrane. The general displacement field of the membrane can be decomposed into $W = \langle W \rangle + \delta W$, in which $\delta W$ corresponds to the scattered wave $P$, according to Eq. (19), with out-of-plane wave vector $k_\perp$. Because the in-plane wave vector, $k_\parallel$ will usually be much greater than $k_1$, the out-of-plane wave vector, $k_\perp = \sqrt{k_1^2 - k_\parallel^2}$, is then imaginary, and the scattered wave $P$ is eventually an evanescent wave. On the other hand, $\langle W \rangle$ is related to a piston-like motion of the MAM, according to Eq. (17), and will contribute to the far field transmission. At the dip frequency with $\langle W \rangle = 0$, $\rho_{\text{eff}}$ will go to the infinity according to the definition of the effective mass density. In conclusion, the frequency range of the negative effective mass density of the MAM corresponds to that with the negative slope of the transmission coefficient.

## IV. RESULTS AND DISCUSSIONS

The proposed analytical model will be further applied for analyzing the coupled vibroacoustic behavior of the MAM. Attention will be paid on microstructural parameter effects such as weight, size, number, shape and location of the attached masses, pretension and thickness of the membrane on acoustic properties of the MAM. Specially, transmission peak and dip frequencies of the MAM with central or eccentric attached masses and two semicircular masses will be quantitatively investigated, respectively.

**A. A MAM with a central mass**

As we know, weight of the attached mass of the MAM is a key parameter of selecting the resonant frequencies for low-frequency acoustic transmission applications. Figure 4 shows effects of different weights of the central mass on the two peak frequencies and one dip frequency of the MAM. The dip frequency can be determined from the current model by setting the average displacement within the MAM to be $\langle W \rangle = 0$. In the figure, the material properties of the MAM are the same as listed in Table 1 and only the weight of the mass is changed. It can be found that the first peak and dip frequency will be decreased significantly as the weight of the mass increases from 50 mg to 300 mg, while the second peak frequency is not sensitive to the weight change of mass. It is understandable because the first eigenmode (peak frequency) is caused by the translational motion of the mass, and the first resonant frequency is inversely proportional to the square root of the weight of the mass. On the other hand, the second eigenmode (peak frequency) depends strongly on the motion of the membrane, and the motionless mass effect could be very small.

The mass geometry effects on the two peak frequencies and one dip frequency of the MAM are illustrated in Fig. 5. In the figure, the material properties of the MAM are the same as listed in Table 1 and only the radius of the mass with the same weight is changed. As shown in the figure, when the size (radius) of the mass is reduced, the first peak frequency remains almost unchanged, and both the second peak frequency and the dip frequency are significantly decreased. For example, when the radius of the mass is reduced from 3mm to 1.5mm, the second peak frequency is decreased from 965Hz to 780Hz, the dip frequency is decreased from 265Hz to 170Hz, and the first peak frequency is almost unchanged from 145Hz to 120Hz. Those findings could be used for the MAM design of targeting different transmission peak and/or dip frequencies.

Figure 6 shows the membrane's pretension effect on the two peak frequencies and one dip frequency of the MAM. The material properties of the MAM are the same as listed in Table 1 and only the pretension of the membrane is changed. It is noticed that the first peak frequency and the dip frequency are not sensitive to the change of the pre-stress in the membrane, but the second peak frequency does increase a lot with the change of the pre-stress. This is because the increase of the pretension can lead to the increase of elastic wave speed of the membrane, and therefore the second peak frequency increases as expected. Finally, the membrane thickness effect on the two peak frequencies and one dip frequency of the MAM is studied in Fig. 7. We found that the increase of the thickness of the membrane can only reduce the second peak frequency, neither the first peak frequency nor the dip frequency will be affected significantly.

## B. A MAM with an eccentric mass

Figure 8 shows acoustic transmissions of the MAM with an eccentric attached mass with the eccentricity $d = 6, 4, 2$ mm, respectively. In the figure, the material properties of the MAM are the same as listed in Table 1. It is interesting to notice that new transmission peak and dip frequencies are observed, compared with the MAM with a central mass. Based on the eigenmodes shown in Fig. 2(b), at the first peak frequency, the eccentric mass has not only the strong translational motion, but also the rotational motion. The new peak frequency is caused by the flapping of the mass in conjunction with a weak translational motion. The third peak frequency is mainly caused by the vibration of the membrane in conjunction with small translational and rotational motions of the mass. Among these three peak frequencies, two dip frequencies can be found. Figure 9 shows the trend of the three peak frequencies with the change of the mass eccentricity. It is evident that the first and second peak frequencies will increase with the increase of the mass eccentricity, and the third peak frequency will significantly decrease with the increase of the mass eccentricity. For example, the third peak frequency is reduced from 970Hz to 650Hz when the dimensionless mass eccentricity is changed from $d/R = 0$ to 0.6.

## C. A MAM with two eccentric masses

The current model can be easily extended to calculate the sound transmission of the MAM with multiple attached masses by rewriting governing equations of Eqs. (7) - (9) for the multiple masses and applying appropriate displacement continuity of Eq. (10) on all collocation points. Considering a MAM with two symmetrically attached eccentric semicircular masses, shown in Fig. 10, the first three eigenmodes of the MAM are demonstrated in Fig. 11(a), (b) and (c) at frequencies 215.5 Hz, 518.8 Hz and 956.4 Hz, respectively. In the figure, the membrane's material properties, radius and pretention are the same as listed in Table 1. The radius and the weight of each semicircular mass are 4.5 mm and 200 mg. It can be found that the first eigenmode is caused by translational and rotational motions of semicircular masses in the same phase with the membrane, while the second eigenmode is mainly caused by the rotational motion of the masses. The third eigenmode is due to the strong motion of the membrane in the regions between the two masses. Figure 12 shows acoustic transmissions of this MAM with the eccentricity $d = 3.5$, 4.0, 4.5 mm, respectively. In the figure, the membrane's material properties and pretention are the same as listed in Table 1. The radius and the weight of each semicircular mass are 4.5 mm and 200 mg. As expected, three transmission peaks and two transmission dips are observed, which are caused by three resonant and two anti-resonant (with $\langle W \rangle = 0$) modes, respectively. In the model, we do not consider the membrane's dissipation effects, therefore, transmissions can reach 100% at resonant frequencies. The trend of three peak frequencies with the change of the mass eccentricity is shown in Fig. 13. It is clear that the mass eccentricity could provide an additional design space for tuning acoustic transmission peaks and dips of the MAM at different frequencies.

## V. CONCLUSIONS

In this paper, an analytical model is developed to capture the coupled vibroacoustic behavior of the MAM. Different from existing models, the proposed model can provide highly precise analytical solutions of vibroacoustic problems of the MAM, in which the finite mass effects on the deformation of the membrane can be properly captured. The accuracy of the model is verified through the comparison with the FE method. In particular, the microstructure effects such as weight, size, number, shape and eccentricity of masses, pretension and thickness of the membrane on the resulting peak and dip frequencies of the MAM are quantitatively investigated. It is found that the mass eccentricity of the MAM can provide additional transmission peak and dip frequencies.

**ACKNOWLEDGMENTS**

The authors would like to thank Dr. Ping Sheng from Hong Kong University of Science and Technology and Dr. Jun Mei from South China University of Technology, whose comments and discussions improved this work significantly. This work was supported in part by the Air Force Office of Scientific Research under Grant No. AF 9550-10-0061 with Program Manager Dr. Byung-Lip (Les) Lee and by National Natural Science Foundation of China under Grants 11221202, 11290153 and 11172038.

**Tables**

Tab. 1 Material Properties and Parameters of Membrane and Mass

|  | Membrane | Mass |
|---|---|---|
| Mass Density (kg/m$^3$) | 980 | 7800 |
| Young's modulus (Pa) | $2\times10^5$ | $2\times10^{11}$ |
| Poisson ratio | 0.49 | 0.33 |
| Thickness (mm) | 0.28 | - |
| Total Weight (mg) | - | 300 |
| Radius (mm) | 10 | 3 |
| Pretension (N/m) | 51.2 | 0 |

Tab. 2 Comparison between eigenfrequencies by analytical model and FE simulation

|  | Analytical | FEA |
|---|---|---|
| 1-st Mode ($d = 0$ mm) | 145.3 Hz | 145.9 Hz |
| 2-nd Mode ($d = 0$ mm) | 971.6 Hz | 985.5 Hz |
| 3-rd Mode ($d = 0$ mm) | 1948.5 Hz | 1981.8 Hz |
| 1-st Mode ($d = 6$ mm) | 178.6 Hz | 177.6 Hz |
| 2-nd Mode ($d = 6$ mm) | 400.6 Hz | 414.0 Hz |
| 3-rd Mode ($d = 6$ mm) | 646.5 Hz | 653.8 Hz |

*Figure Captions:*

FIG. 1. (a) MAM attached with an eccentric mass. (b) MAM subjected to a normal acoustic loading in a cylindrical tube.

FIG. 2. (a) Mode shapes of MAM with a central mass. First row: analytical solution. Second row: Finite element solution. (b) Mode shapes of MAM with an ecccentric mass. First row: analytical solution. Second row: Finite element solution.

FIG. 3. Validation of transmissions of MAM with a central mass and its effective dynamic mass density.

FIG. 4. Effects of mass's weight to peak and dip frequencies of MAM with a central mass.

FIG. 5. Effects of mass's radius to peak and dip frequencies of MAM with a central mass.

FIG. 6. Effects of membrane's pretension to peak and dip frequencies of MAM with a central mass.

FIG. 7. Effects of membrane's thickness to peak and dip frequencies of MAM with a central mass.

FIG. 8. Transmissions of MAM with an eccentric mass with different eccentricities.

FIG. 9. Effects of mass's eccentricity to peak frequencies of MAM with an eccentric mass.

FIG. 10. MAM symmetrically attached with two semicircular masses.

FIG. 11. Mode shapes of MAM with two symmetrically attached semicircular masses: (a) first mode; (b) second mode; (c) third mode.

FIG. 12. Transmissions of MAM with two symmetrically attached semicircular masses with different eccentricities.

FIG. 13. Effects of masses' eccentricities to peak frequencies of MAM with two symmetrically attached semicircular masses.

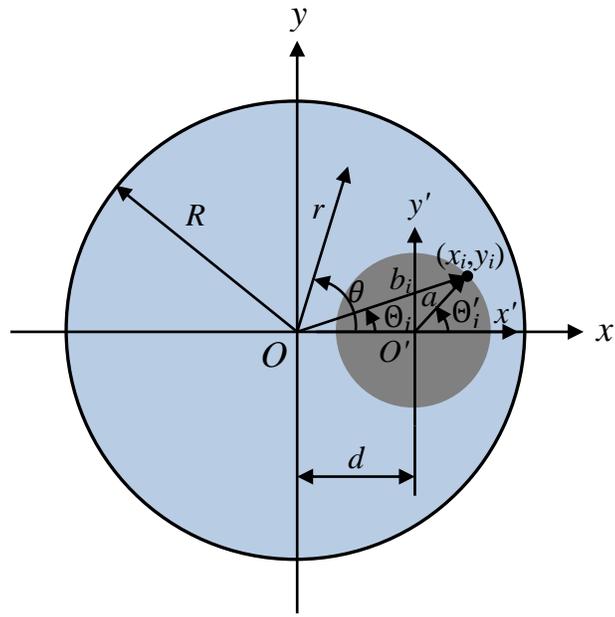

(a)

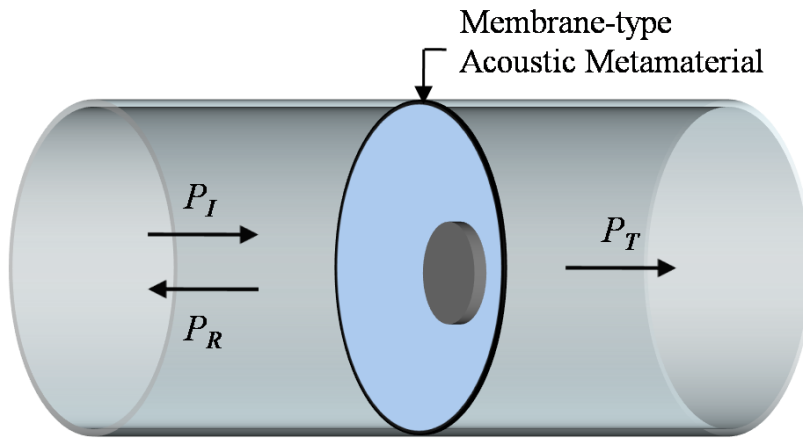

(b)

FIG. 1

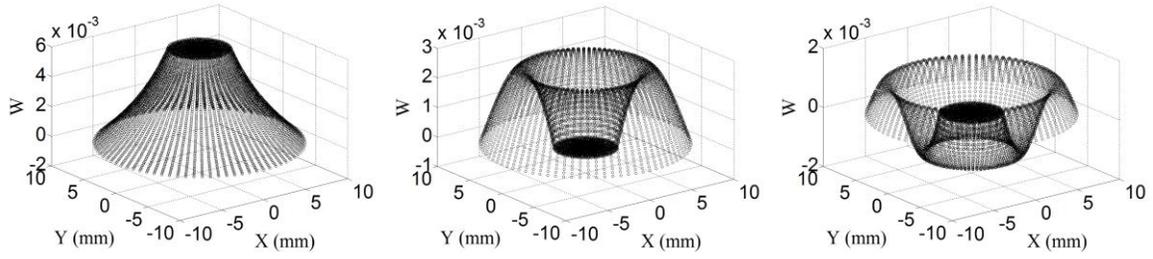

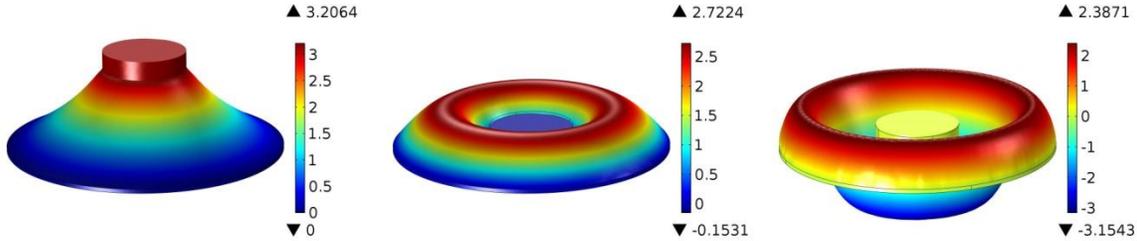

(a)

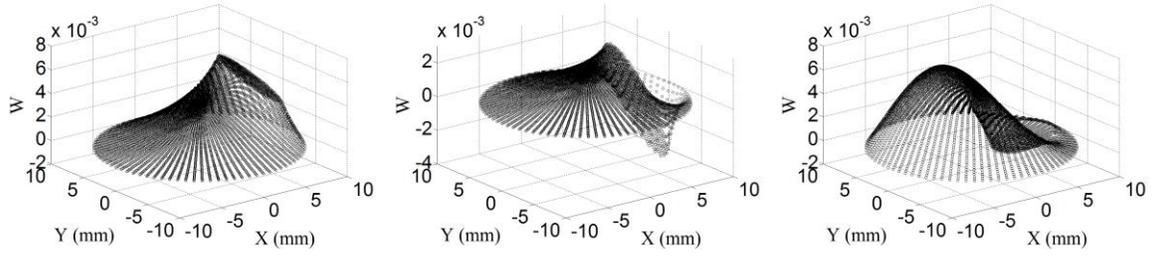

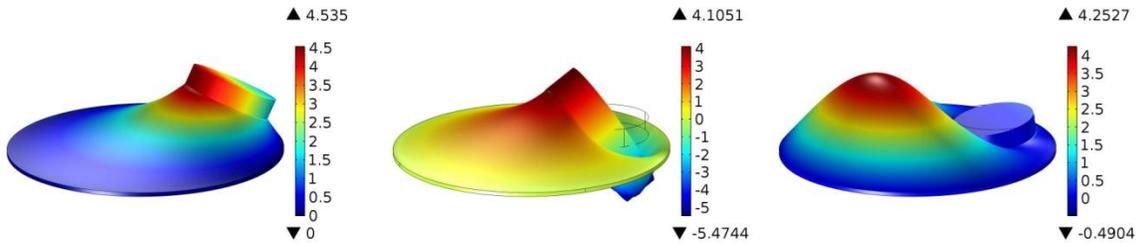

(b)

FIG. 2

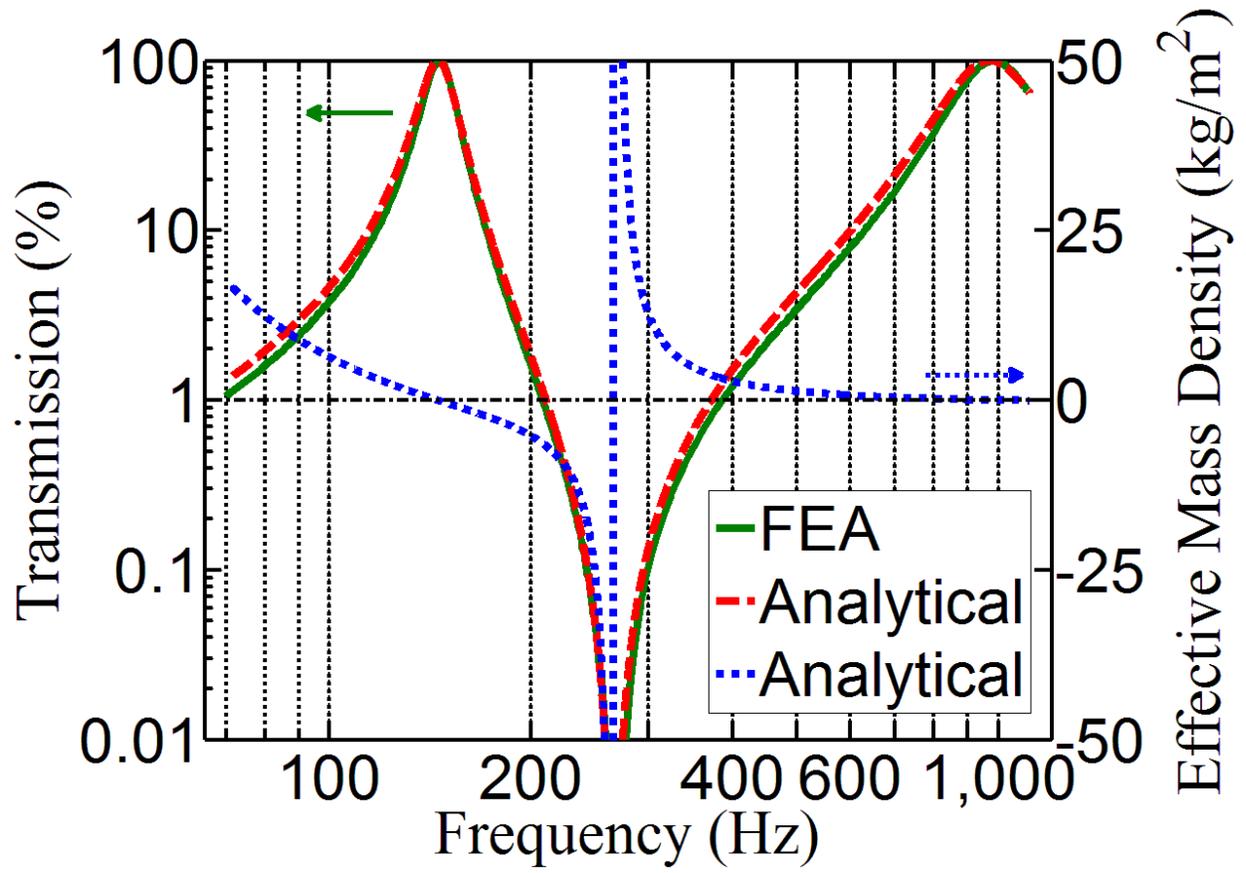

FIG. 3

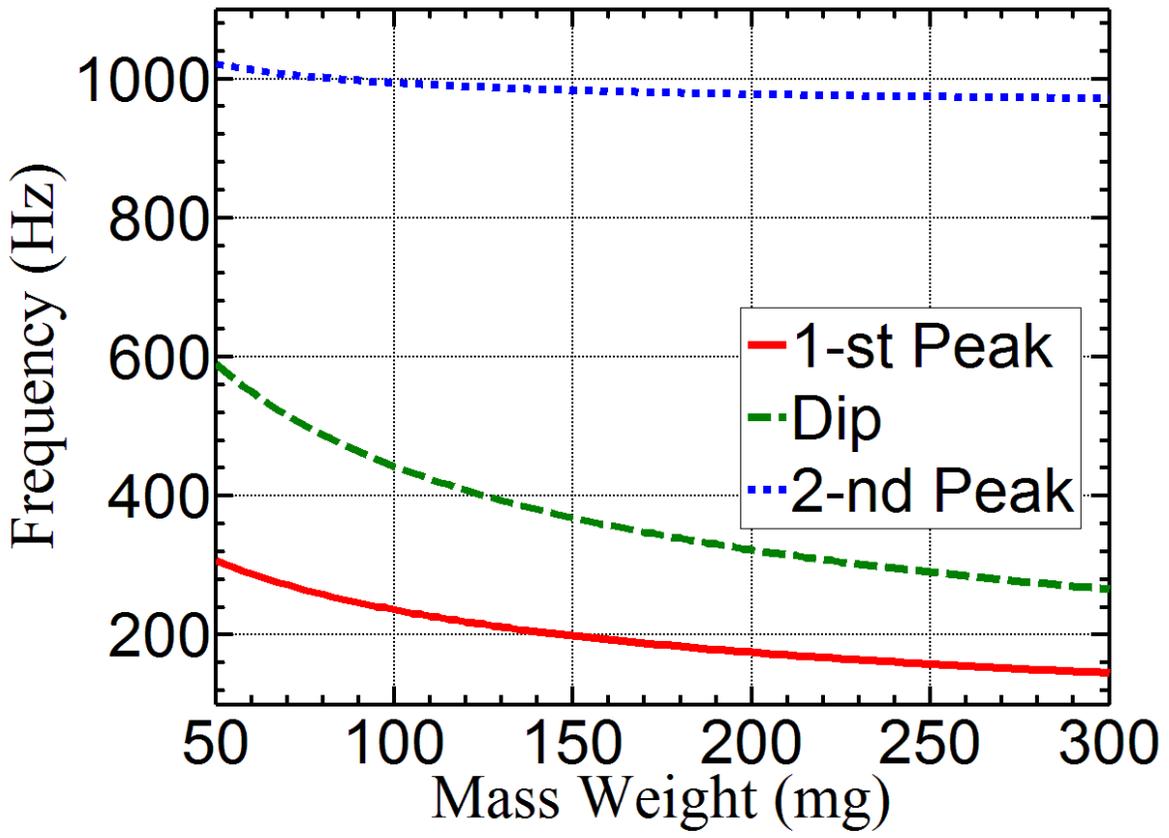

FIG. 4

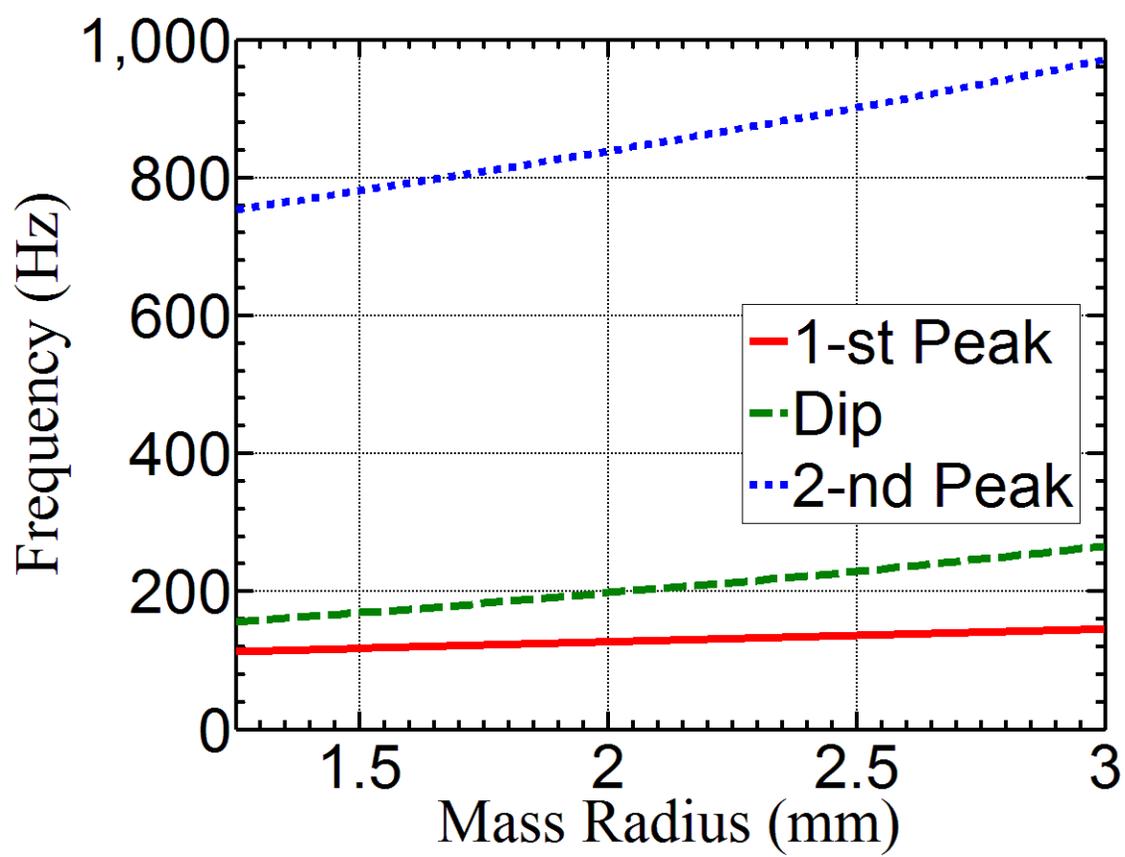

FIG. 5

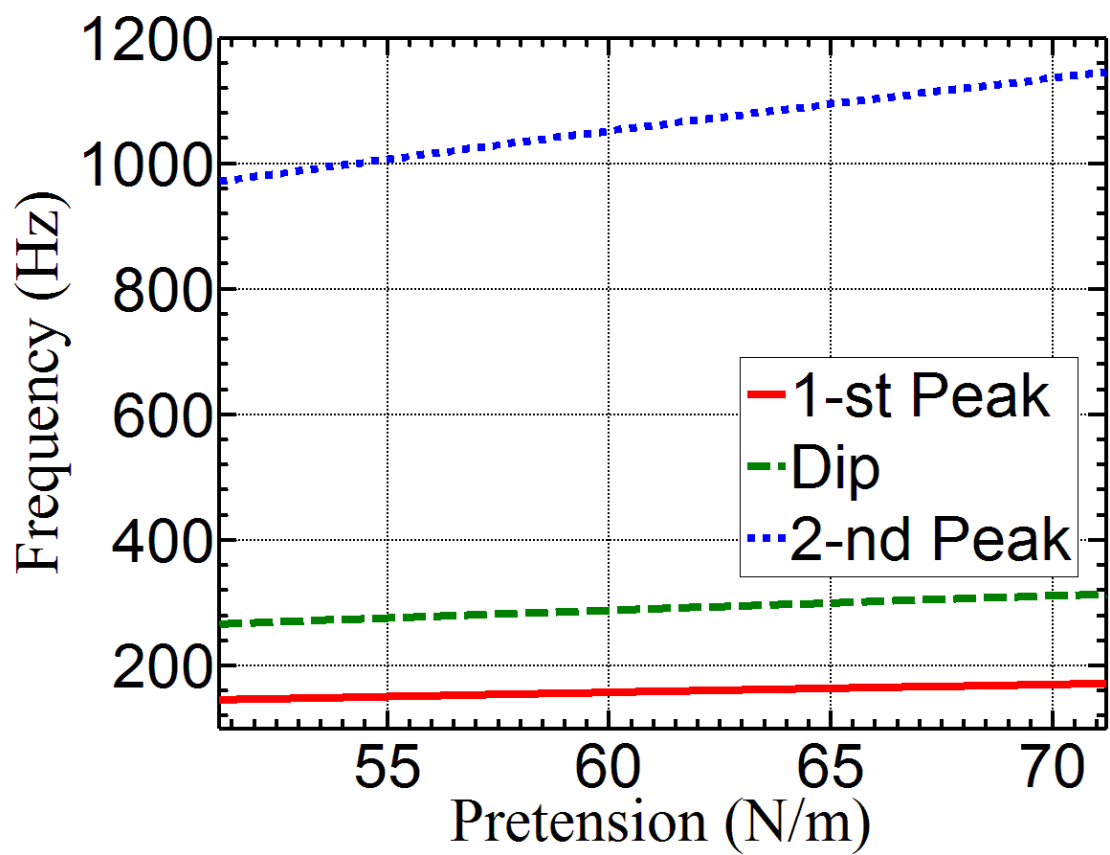

FIG. 6

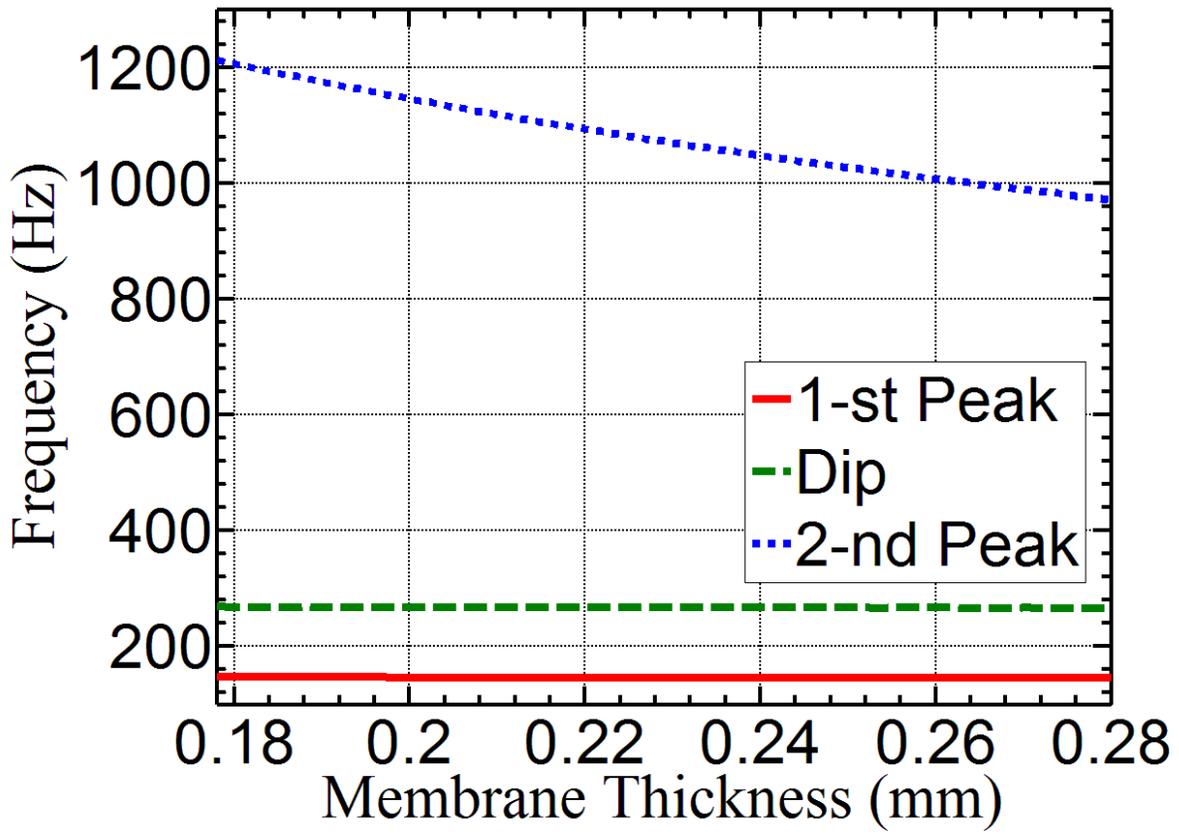

FIG. 7

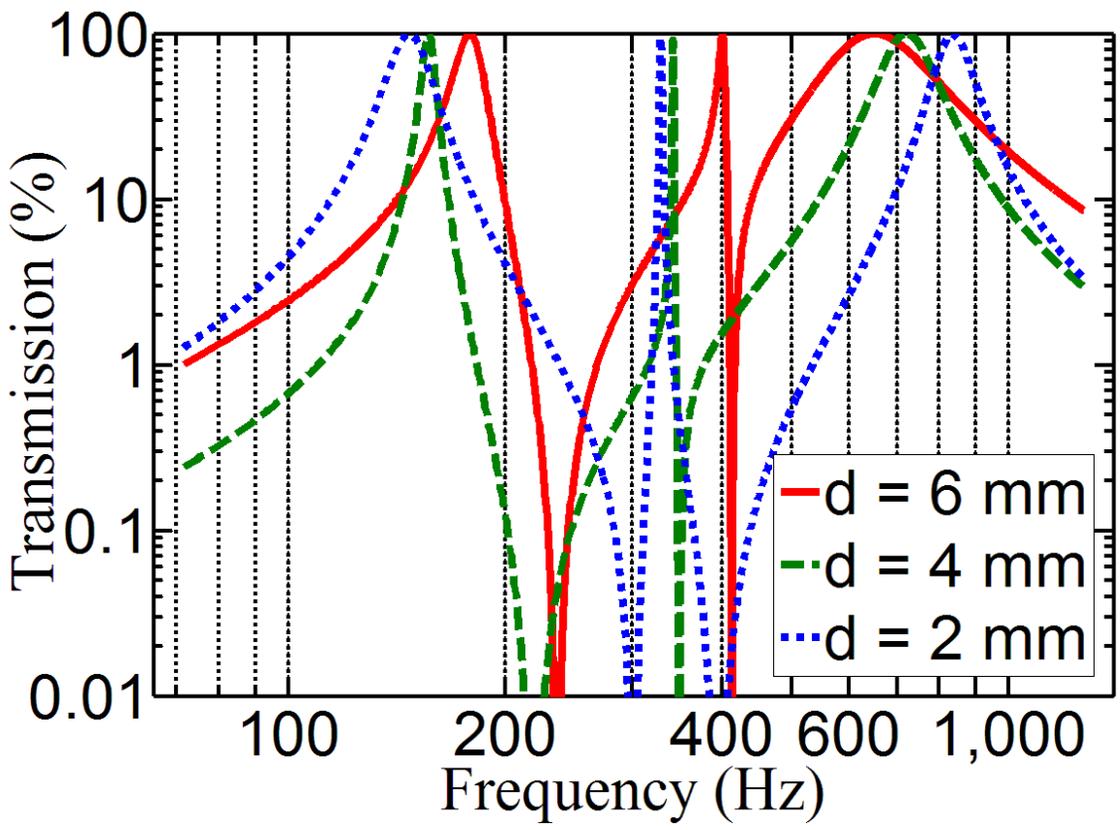

FIG. 8

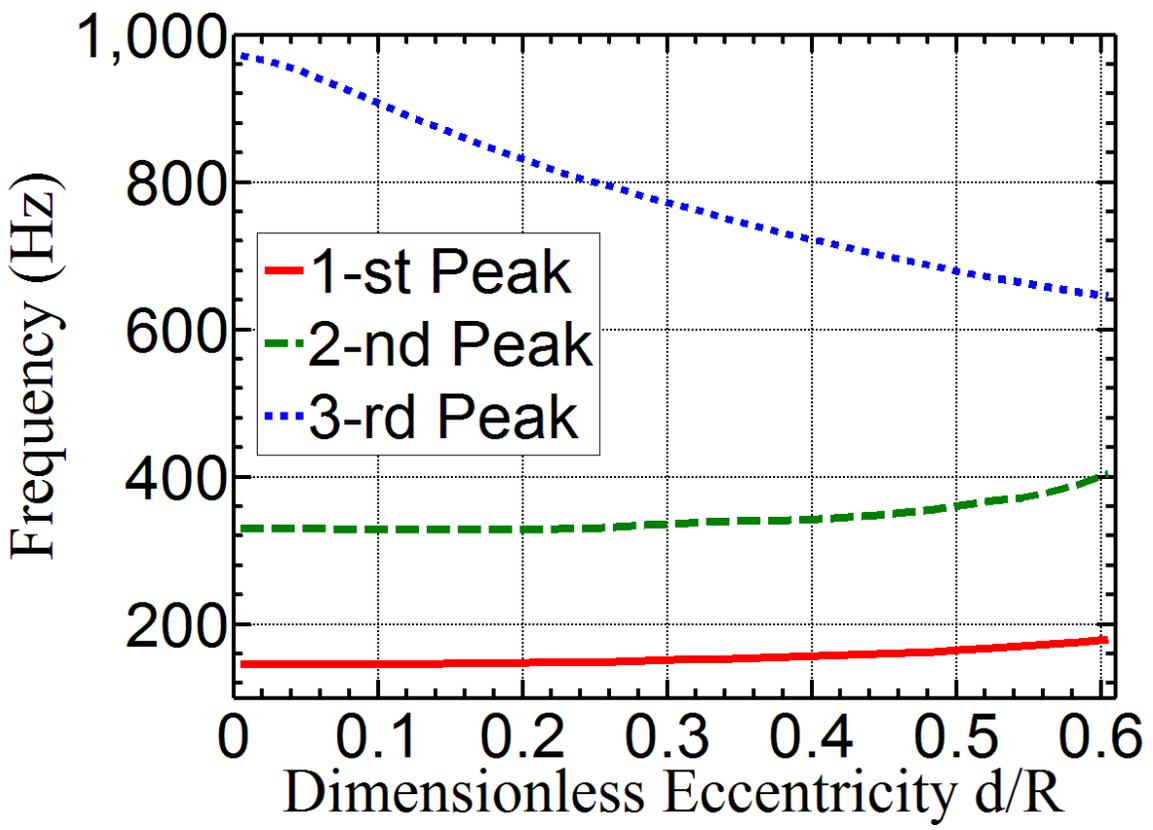

FIG. 9

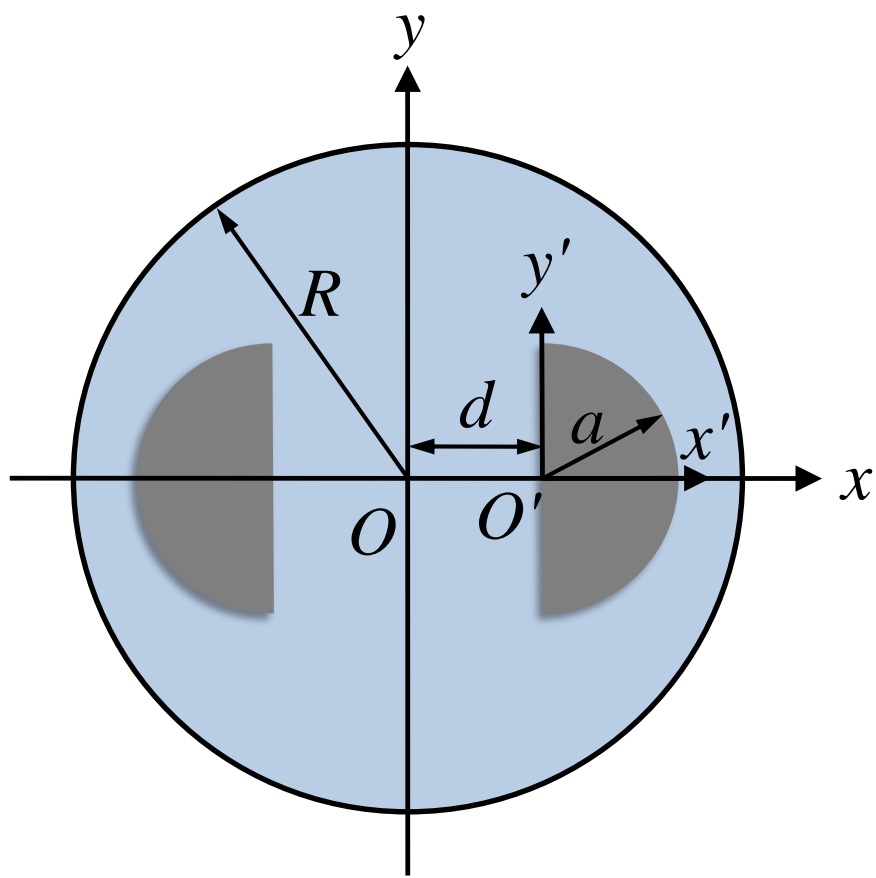

FIG. 10

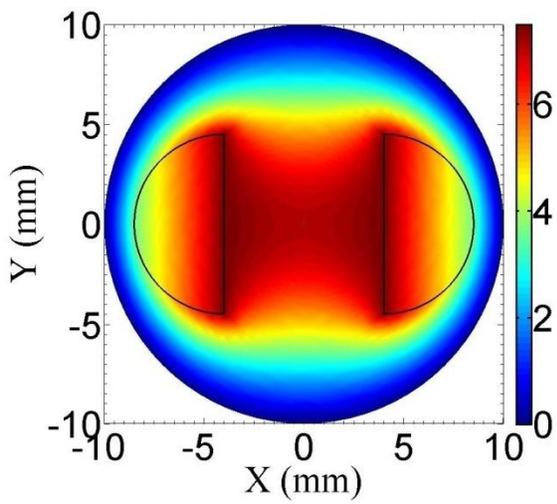

(a)

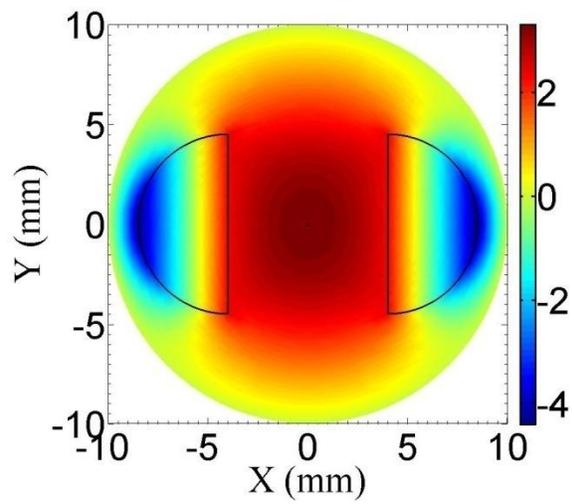

(b)

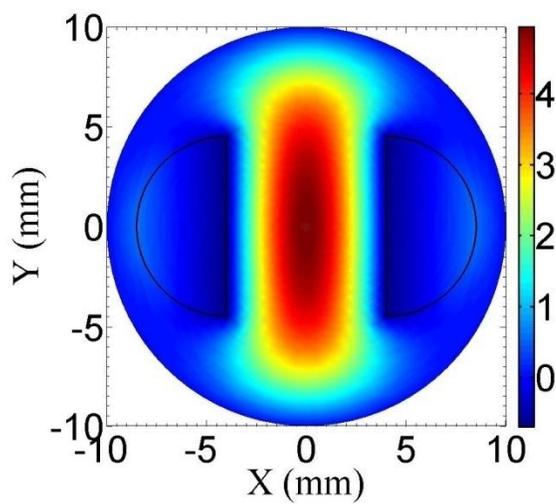

(c)

FIG. 11

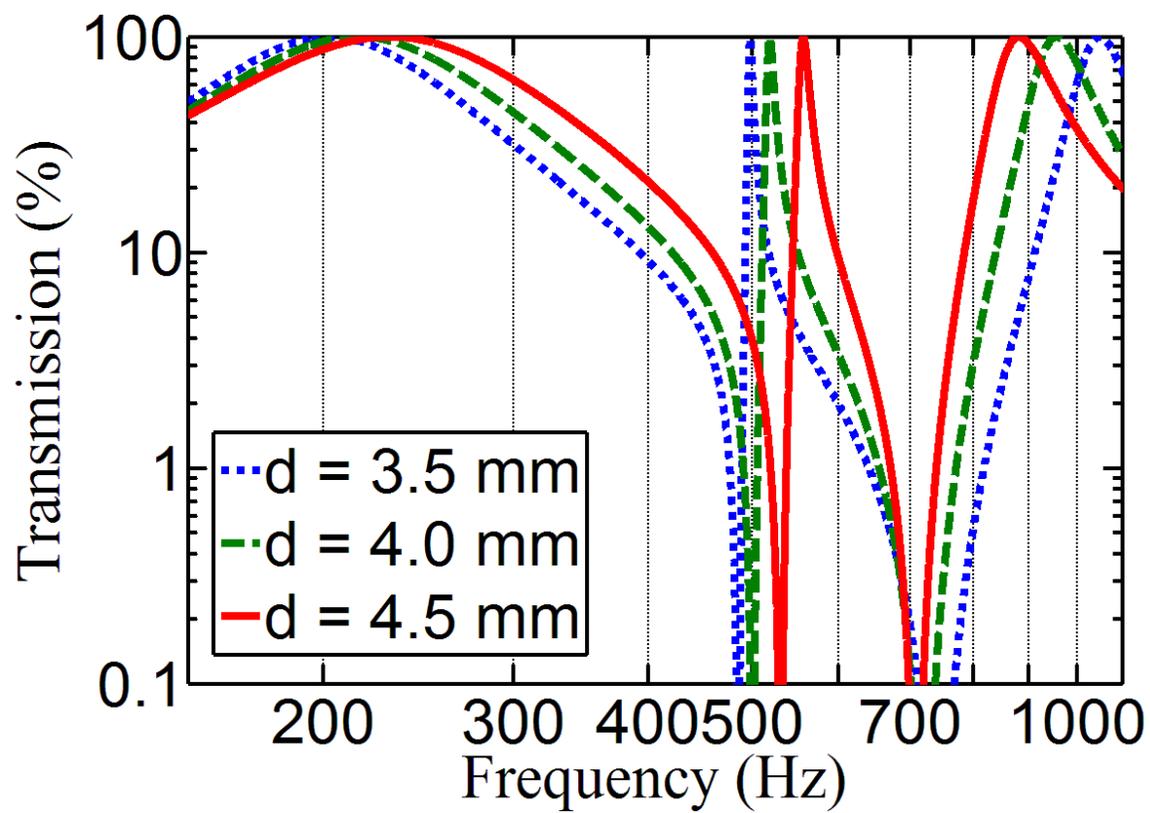

FIG. 12

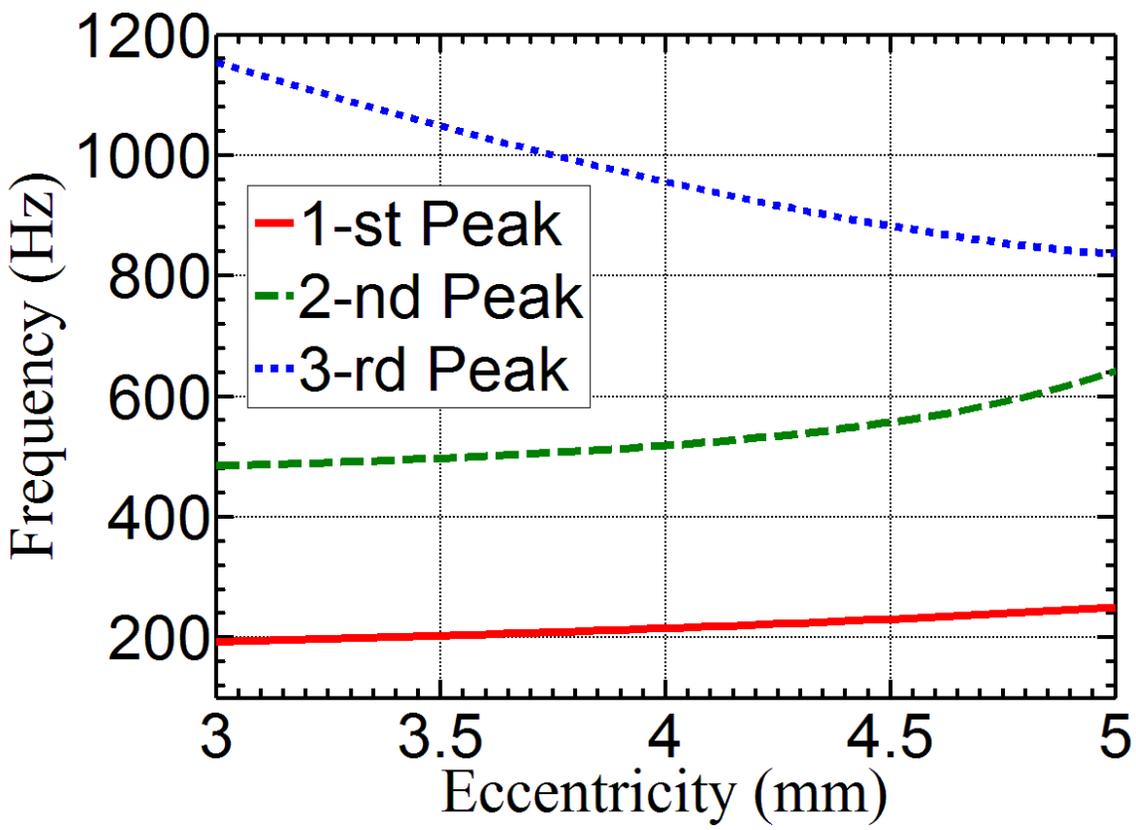

FIG. 13